\documentclass[%
 aip,
 amsmath,amssymb,
 reprint,%
]{revtex4-1}

\usepackage{graphicx}
\usepackage{dcolumn}
\usepackage{bm}

\usepackage[utf8]{inputenc}
\usepackage[T1]{fontenc}
\usepackage{mathptmx}
\usepackage{etoolbox}

\usepackage{amsmath,fullpage,hyperref}
\usepackage{array}
\usepackage{textcomp}
\usepackage{amssymb}
\usepackage{amsfonts}
\usepackage{soul}

\usepackage{xcolor}

\newcommand\dd{\mathrm{d}}

\providecommand\bnabla{\bm{\nabla}}

\renewcommand{\vec}[1]{\ensuremath{\mathbf{#1}}} 
 
\providecommand\bnabla{\gvec{\nabla}}

\newcommand{\aderv}[2]{\dfrac{\partial {#1}}{\partial {#2}}}

\newcommand{\adervs}[2]{\dfrac{\partial^2 {#1}}{\partial {#2}^2}}

\newcommand{\equa}[1]{Eq.~(\ref{#1})}
\newcommand{\fig}[1]{Fig.~\ref{fig#1}}

\newcommand{\equass}[2]{Eqs.~(\ref{#1})--(\ref{#2})}
\newcommand{\equasa}[2]{Eqs.~(\ref{#1}){ }and{ }(\ref{#2})}

\newcommand{\eqn}[2]{\begin{gather}
#1
\label{#2}
\end{gather}
}

\usepackage{fullpage,color}

\makeatletter
\def\@email#1#2{%
 \endgroup
 \patchcmd{\titleblock@produce}
  {\frontmatter@RRAPformat}
  {\frontmatter@RRAPformat{\produce@RRAP{*#1\href{mailto:#2}{#2}}}\frontmatter@RRAPformat}
  {}{}
}%
\makeatother
\begin{document}

\preprint{AIP/123-QED}

\title[]{Effects of anisotropy on the transition to absolute instability in a porous medium heated from below}
\author{M. Celli}
 \altaffiliation[corresponding author: ]{michele.celli3@unibo.it}
\author{A. Barletta}%
\affiliation{$\mbox{}^1$Alma Mater Studiorum Universit\`{a} di Bologna, Department of Industrial Engineering, Viale Risorgimento 2, 40136 Bologna, Italy
}%

\date{\today}

\begin{abstract}
The emerging instability of a forced throughflow in a fluid saturated horizontal porous duct of rectangular cross section is investigated. The duct is heated from below by assuming the horizontal boundaries to be at different temperatures. Both the horizontal and the vertical boundaries are impermeable and the basic flow is parallel to such boundaries. The porous medium is anisotropic with different permeabilities in the vertical and horizontal directions. The effect of the anisotropy on the onset of the buoyancy driven modal instability and absolute instability is analysed. The parametric conditions leading to the instability of the basic flow are determined by employing an analytical dispersion relation. 
The different permeabilities in the vertical and horizontal directions come out to play opposite roles in the onset of modal instability and in the transition to absolute instability. It is shown that an increasing vertical permeability has a destabilising effect, while an increasing horizontal permeability has a stabilising effect.
\end{abstract}

\maketitle

\section{Introduction}
{The onset of convection in a horizontal porous layer heated from below has widespread applications either in engineering and geophysics. Such applications regard, for instance, the dynamics of groundwater reservoirs, the diffusion of pollutants in the soil, the ${\rm CO_2}$ sequestration and the thermal insulation of buildings. In several cases, the porous media involved in these applications are anisotropic.}

The study reported in this paper contributes further novel findings to the present knowledge on the buoyancy driven instability in fluid saturated horizontal porous layers heated from below \cite{NiBe17}. This is also well--known as the Darcy--B\'enard problem, or as the Horton--Rogers--Lapwood (HRL) problem. In fact, the first investigations on this topic were carried out by Horton and Rogers \cite{horton1945convection}, and by Lapwood \cite{lapwood1948convection}. These authors considered a porous layer infinitely wide in the horizontal directions and bounded in the vertical direction by two impermeable boundaries held at different temperatures, such that an upward vertical temperature gradient is imposed. By assuming a layer saturated with a motionless fluid,  the threshold conditions for the onset of modal instability were evaluated finding that convective cells may arise when the Darcy--Rayleigh number becomes greater than $4\, \pi^2$. When Prats \cite{prats} further developed the HRL problem by including a basic horizontal throughflow, he found out that the threshold for the onset of modal instability was not affected by the mass flow rate. 

Barletta \cite{barletta2019r} presents a survey of the recent research studies regarding both the modal instability and the transition to the absolute instability for the Prats problem. While the modal instability analysis is focussed on investigating the evolution in time of single Fourier modes perturbing the whole basic state, the absolute instability analysis deals with the evolution of perturbation wavepackets expressed by a Fourier integral over all possible wavenumbers.  The latter type of instability is the one that is typically observed in the laboratory reference frame when a localised disturbance spreads in space and travels in the direction of the basic flow.  For this  reference frame, amplified perturbations can be detected only when they are not convected downstream by the basic throughflow. By illustrating the transition to absolute instability for the Prats problem, \citet{barletta2019r} points out that the threshold value of the governing parameter, {\em i.e.} the Rayleigh number, is a monotonically increasing function of the basic flow rate, parametrised through the P\'eclet number. By increasing the flow rate, or the P\'eclet number, the unstably growing perturbation is more likely to be convected downstream, so that the basic flow becomes more stable.

{An interesting reformulation of the HRL problem accounting for the anisotropy of the porous medium has been performed in several papers \cite{CC,MK,qin1994convective,storesletten1998,PU01,AR01,AR06,SR}. These authors carried out} the modal stability analysis assuming the thermophysical properties of the porous media to have constant although different values along the three Cartesian directions. The anisotropy effect on the convection in porous media has been investigated also by \citet{rees2002convective,ennis2005onset,nield2007effects,de2016influence}.

The analysis presented in this paper is focussed on the investigation of both modal and absolute instabilities for the Prats problem where an anisotropic porous duct, with a rectangular cross section, is considered. In particular, the porous medium is characterised by two different uniform values of permeability: a value for the permeability in the horizontal directions and a value for the permeability in the vertical  direction. This work fills a gap in the literature since, to the best of authors' knowledge, the absolute instability analysis for the anisotropic Prats' problem in a rectangular duct has not been investigated to date.

\section{Mathematical model}
\begin{figure*}
\centering
\includegraphics[width=12cm]{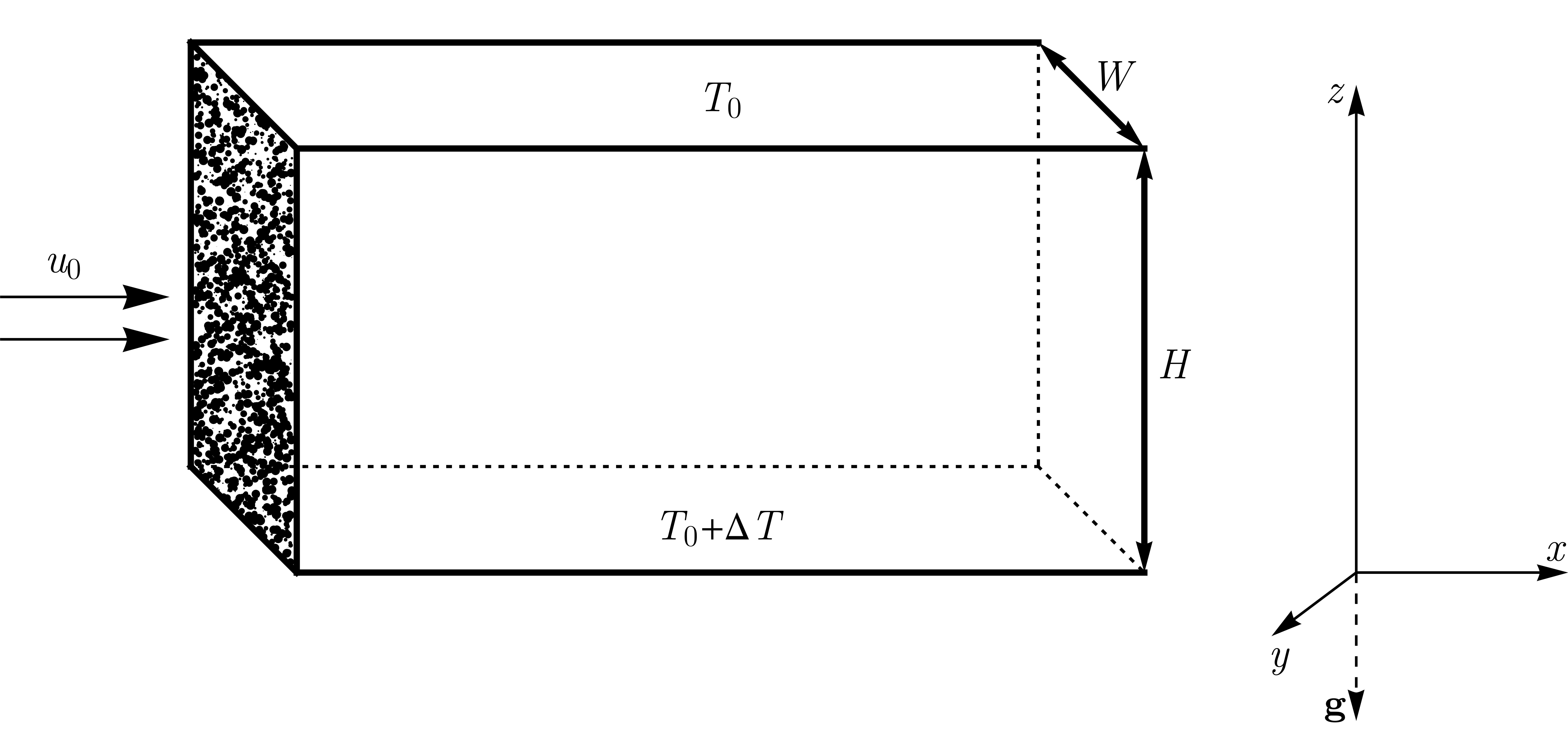}
\caption{Sketch of the porous duct geometry and boundary conditions.}
\label{fig1}
\end{figure*}
A fluid saturated horizontal porous duct of rectangular cross section with height $H$ and width $W$ is considered. This duct is assumed to be unbounded in the $x$ direction, while it is bounded by impermeable walls in the $y$ and $z$ directions. The duct is heated from below with the horizontal boundaries having different temperatures: the boundary at $z=0$ is held at temperature $T_0 +\Delta T$, with $\Delta T > 0$, while the boundary at $z=H$ is held at temperature $T_0$, see \fig{1}.

{The momentum transport is formulated by employing Darcy's law together with the Oberbeck-Boussinesq approximation. The energy transport is modelled by assuming a negligible viscous dissipation and local thermal equilibrium between the fluid and solid phases. A scaling is used for the non--dimensional formulation,
\eqn{
\dfrac{\vec x}{H} \to \vec x , \quad  \dfrac{\chi}{ H^2} t \to t, \quad  \dfrac{H}{\chi} \vec u \to \vec u ,\nonumber \\
   \dfrac{K_z}{\chi \, \mu}p\to p, \quad \dfrac{T-T_0}{\Delta T} \to T.
}{3}}
Thus, the governing balance equations, in dimensionless form, are given by 
\eqn{
\bnabla \cdot \vec u =0, \nonumber\\
 u = -a\aderv{p}{x} , \nonumber\\
 v =-a\,\aderv{p}{y} ,\nonumber\\
 a \, w =-a\, \aderv{p}{z}+R\, T , \nonumber\\
 \sigma \aderv{T}{t}+ \vec u \cdot  \bnabla\,  T=  \nabla ^2  T,
 }{1ant}
 with the boundary conditions
\eqn{
 y=0,s: \qquad  v=0 , \quad \aderv{T}{y}=0 ,\ \nonumber\\
z=0: \qquad  w=0 , \quad T=1 ,  \nonumber\\
z=1: \qquad  w=0 , \quad T=0.
}{1}
In \equass{3}{1}, $p$ denotes the local difference between the pressure and the hydrostatic pressure (hereafter, $p$ is just called pressure for the sake of brevity), $(x,y,z)$ are the Cartesian components of the position vector $\vec{x}$, $(u,v,w)$ are the Cartesian components of the seepage velocity vector $\vec{u}$, $t$ is the time, $T$ is the temperature, $s=W/H$ is the aspect ratio of the duct and  $\sigma$ is the heat capacity ratio, namely the ratio between the heat capacity per unit volume of the saturated porous medium and the heat capacity per unit volume of the fluid \cite{NiBe17}. Furthermore, the Darcy--Rayleigh number $R$ is defined as
\eqn{
R=\frac{\rho \, g \, \beta \, K_h \, H \, \Delta T}{\mu \, \chi },
}{2R}
where  $\rho$ is the fluid density at the reference temperature $T_0$, $g$ is the modulus of the gravitational acceleration $\vec{g}$, $\beta$ is the thermal expansion coefficient of the fluid, $\mu$ is the dynamic viscosity of the fluid and $\chi$ is the average thermal diffusivity of the fluid saturated porous medium. 

We defined $R$ using the horizontal permeability $K_h$ instead of the vertical, $K_z$, because the definition (\ref{2R}) simplifies the interpretation of the results obtained by the forthcoming absolute instability analysis. We mention that \citet{storesletten1998} defines the Darcy--Rayleigh number differently by employing $K_z$ instead of $K_h$. 

{The quantity $\chi$ is defined as the ratio between the average thermal conductivity of the fluid saturated porous medium and the volumetric heat capacity of the fluid, i.e., the product $\rho \, c$, where $c$ is the specific heat of the fluid. Following the reasoning presented by \citet{BV} and by \citet{VK}, the thermal dispersion effects have been implicitly taken into account in the average thermal diffusivity $\chi$.}

The parameter $a$ is the ratio between the permeability of the porous medium in the horizontal directions, $K_h$, and the permeability in the vertical direction, $K_z$, namely
\eqn{
a=\frac{K_h}{K_z}.
}{2a}
Since Darcy's law is employed to model the momentum transport, the values of $K_h$ and $K_z$ have to be small enough. This constraint implies that  the limiting case $a\to 0$ has to be obtained by assuming $K_h\ll K_z$ with $K_z$ yet sufficiently small in compliance with Darcy's law. 
\section{Basic state}
A steady state solution of \equasa{1ant}{1} is given by a uniform velocity profile, with value $Pe$, in the $x$ direction and a uniform negative temperature gradient in the $z$ direction, namely
\eqn{
u_b=Pe  , \quad v_b=w_b=0 , \quad  T_b=1-z,\nonumber\\
\aderv{p_b}{x}=-\frac{Pe}{a} , \quad \aderv{p_b}{y}= 0 , \quad  \aderv{p_b}{z}=\frac{R}{a} \, (1-z) ,
}{4}
where the subscript ``$b$'' indicates that \equa{4} defines the basic state. The basic state (\ref{4}) yields a pure conduction regime with the heat flux parallel to the $z$ direction. In \equa{4}, $Pe$ is the P\'eclet number defined as 
\eqn{
Pe=\frac{u_0 \, H}{\chi},
}{Pe}
and $u_0$ is the dimensional basic velocity (see \fig{1}).
\section{Linear stability analysis}
Equations~(\ref{1ant}) and (\ref{1}) may be conveniently rewritten according to a pressure--temperature formulation, namely
\eqn{
 a\adervs{p}{x}+a\adervs{p}{y}+\adervs{p}{z}- \frac{R}{a} \;  \aderv{T}{z} =0,\nonumber\\
 \sigma \aderv{T}{t}-a\aderv{p}{x}\aderv{T}{x}-a\aderv{p}{y}\aderv{T}{y}\nonumber\\
\hspace{2cm}+ \left( \frac{R}{a} \, T-\aderv{p}{z}\right)\aderv{T}{z}=  \nabla ^2  T,\nonumber\\
 y=0,s: \qquad \aderv{p}{y}=0 , \quad \aderv{T}{y}=0 ,\  \ \nonumber\\
 z=0: \qquad \aderv{p}{z}=\frac{R}{a} , \quad T=1 ,  \nonumber\\
 z=1: \qquad \aderv{p}{z}=0 , \quad T=0,\
}{5}
where the impermeability condition in \equa{1} have been reformulated in terms of the pressure by employing the modified Darcy's law, \equa{1ant}. The stability of the basic state (\ref{4}) is analysed by employing small--amplitude perturbations, {\em i.e.},
\eqn{
 p(x,y,z,t)=p_b(x,z) + \varepsilon \, P(x,y,z,t), \nonumber\\
 T( x,y,z,t)=T_b (z)+ \varepsilon \,  \Theta(x,y,z,t),
}{6}
where $\varepsilon \ll 1$ is a positive perturbation parameter.
The linearisation of the governing equations is carried out by substituting \equa{6} into \equa{5} and by neglecting terms $O(\varepsilon^2)$. We thus obtain the governing equations and boundary conditions for the disturbances, namely
\eqn{ 
a\adervs{P}{x}+a\adervs{P}{y}+\adervs{P}{z}- \frac{R}{a} \,  \aderv{\Theta}{z} =0, \nonumber\\
\sigma  \aderv{\Theta}{t}+Pe \aderv{\Theta}{x}-\frac{R}{a} \, \Theta+ \aderv{P}{z} \nonumber\\
\hspace{2cm}=\adervs{\Theta}{x}+\adervs{\Theta}{y}+\adervs{\Theta}{z},\nonumber\\
y=0,s: \qquad \aderv{P}{y}=0 , \quad \aderv{\Theta}{y}=0 ,  \nonumber\\
z=0,1: \qquad \aderv{P}{z}=0 , \quad \Theta=0 .\ \
}{7}
Let us now employ the Fourier transforms
\eqn{
 \tilde P(k,y,z,t) = \frac{1}{\sqrt{2 \pi}} \int_{-\infty}^{\infty} P(x,y,z,t)\, e^{-i \, k x}\, \dd x, \nonumber\\
 \tilde  \Theta(k,y,z,t) = \frac{1}{\sqrt{2 \pi}} \int_{-\infty}^{\infty} \Theta(x,y,z,t)\, e^{-i\, k x}\, \dd x,
}{8}
 in \equa{7}. Then, we obtain a problem where the dependence on the $x$ coordinate is superseded by a parametric dependence on the wavenumber $k$, namely
\eqn{ 
a\adervs{ \tilde P}{y}+\adervs{\tilde P}{z}-a\,k^2\,\tilde P-\frac{R}{a} \aderv{\tilde \Theta}{z} =0, \nonumber\\
\sigma  \aderv{\tilde \Theta}{t}+\left(k^2- \frac{R}{a}+i \, k \,{Pe}\right)\tilde \Theta+\aderv{\tilde P}{z}\nonumber \\
\hspace{2cm}=\adervs{\tilde \Theta}{y}+\adervs{\tilde \Theta}{z},\nonumber\\
y=0,s: \qquad \aderv{\tilde P}{y}=0 , \quad \aderv{\tilde \Theta}{y}=0 ,  \nonumber\\
z=0,1: \qquad \aderv{\tilde P}{z}=0 , \quad \tilde \Theta=0 .\ \
}{9}
The inverse Fourier transforms are reported here for convenience,
\eqn{
 P(x,y,z,t)= \frac{1}{\sqrt{2 \pi}} \int_{-\infty}^{\infty}  \tilde P(k,y,z,t)\, e^{i \, k x}\, \dd k, \nonumber\\
 \Theta(x,y,z,t) = \frac{1}{\sqrt{2 \pi}} \int_{-\infty}^{\infty} \tilde  \Theta(k,y,z,t)\, e^{i\, k x}\, \dd k.
}{8b}
Double Fourier series are employed to formulate the dependence on both $y$ and $z$ for $(\tilde P, \tilde \Theta)$, namely
\eqn{
 \tilde P= \sum_{n=0}^{\infty}\,\sum_{m=1}^{\infty} A \cos\left( \frac{n \pi y}{s} \right) \, \cos\left( m \pi z \right)\, e^{\lambda t} , \nonumber\\
\tilde \Theta= \sum_{n=0}^{\infty} \sum_{m=1}^{\infty} B \cos\left( \frac{n \pi y}{s} \right) \, \sin\left( m \pi z  \right) \, e^{\lambda t} ,
}{10}
where $\lambda$ is a complex parameter: the real part of $\lambda$ is the growth rate of the given Fourier mode while the imaginary part of $\lambda$ coincides with $-\omega$, where $\omega$ is the angular frequency of the given Fourier mode.  With \equa{10},  $(\tilde P, \tilde \Theta)$ identically satisfy the boundary conditions reported in \equa{9}. 
By substituting \equa{10} into \equa{9}, we obtain two algebraic equations which lead us to an explicit dispersion relation,
\eqn{ 
\sigma\,\lambda=\frac{ R \left(k^2+r^2\right)}{a \left(k^2+r^2\right)+\pi ^2 m^2}\nonumber \\
\hspace{2cm}-(k^2+r^2+\pi ^2 m^2+i k Pe),
}{12}
where $r=n \,\pi /s$ is a real non--negative parameter. 
\section{Modal instability}
The determination of the threshold value of $R$ for the onset of modal, or convective, instability is accomplished through the study the long time behaviour of each single Fourier mode. This is the reason for the term ``modal instability''. Thus, the condition of zero growth rate is imposed, \textit{i.e.} $\lambda=-i\, \omega$. One may define a rescaled angular frequency $\xi=\omega- k \, Pe/\sigma$, where $\xi$ is  the angular frequency in the reference frame comoving with the basic flow, so that \equa{12} yields
\eqn{ 
R=\frac{\left( k^2+r^2+\pi ^2 m^2 - i \xi \right)  \left[a \left(k^2+r^2\right)+\pi ^2 m^2\right]}{k^2+r^2}.
}{12tris}
Since $R$ is real, also the right hand side of \equa{12tris} must be real. Then, we conclude that $\xi=0$: the angular frequency in the comoving reference frame of the basic flow is zero and, for this reference frame, the principle of exchange of stabilities holds. Moreover, the Darcy--Rayleigh number simplifies to
\eqn{ 
R=\frac{\left(k^2+r^2+\pi ^2 m^2\right) \left[a \left(k^2+r^2\right)+\pi ^2 m^2\right]}{k^2+r^2}.
}{12bis}
It is worth noting that \equa{12bis} agrees with the neutral stability condition reported in the literature \cite{storesletten1998}, when the Darcy--Rayleigh number is defined by employing the vertical permeability $K_z$ instead of the horizontal permeability $K_h$. We mention that \citet{storesletten1998} discusses the stability analysis of the motionless basic state $(Pe=0)$. We emphasise that, exactly as for the Prats' problem in an isotropic medium \cite{prats}, the threshold value for the onset of the modal instability is not affected by the presence of a basic throughflow. This conclusion is easily gathered from \equa{12bis}.

The neutrally stable Darcy--Rayleigh number $R$ in \equa{12bis} is a monotonic increasing function of $m$. Therefore,  for the evaluation of the critical value of $R$, we can set $m$ at its lowest, namely $m=1$. Then, the minimum of the function $R(\zeta)$, where $\zeta=\sqrt{k^2 + r^2}$, yields the critical values of $R$ and $\zeta$ for the onset of the modal instability, 
\eqn{ 
R_c=\pi ^2  \left(\sqrt{a}+1\right)^2, \quad \zeta_c=\frac{\pi}{a^{1/4}}.
}{13}
{The critical values reported in \equa{13} coincide with those obtained by \citet{CC} and reported by \citet{NiBe17} by considering a vanishing throughflow.} For the limiting case of an isotropic porous medium, $a=1$, \equa{13} yields the critical values obtained by Prats \cite{prats}, $R_c=4\pi ^2$ and $\zeta_c=\pi$.  The critical values of $R$ as a function of the parameter $a$ are reported in \fig{2}.
%
\begin{figure}
\centering
\includegraphics[width=1\linewidth]{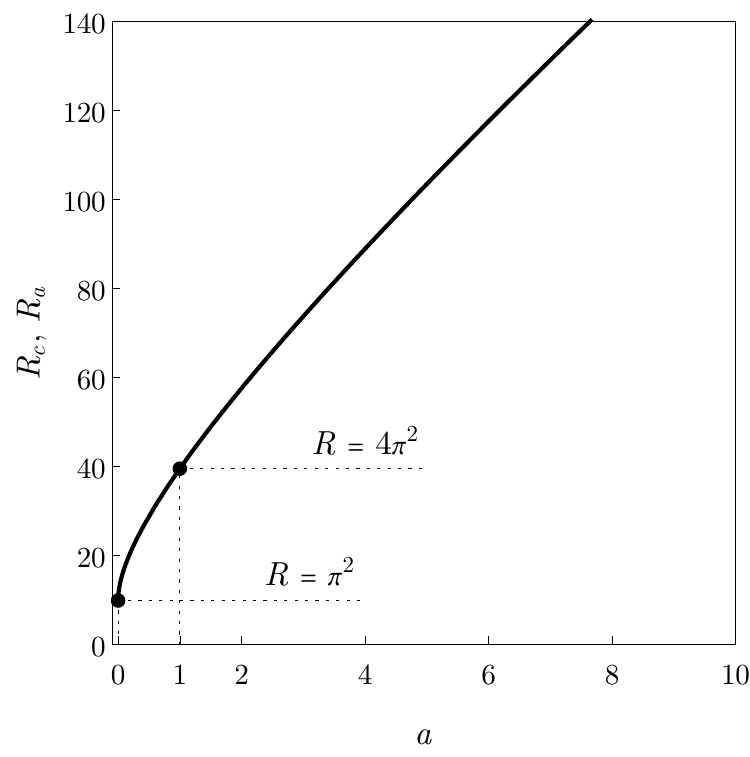}
\caption{Threshold value for the onset of modal instability $R_c$ versus $a$ as coincident with the threshold value for the onset of absolute instability $R_a$ for the limiting case $Pe \to 0$.}
\label{fig2}
\end{figure}

\begin{figure}
\centering
\includegraphics[width=1\linewidth]{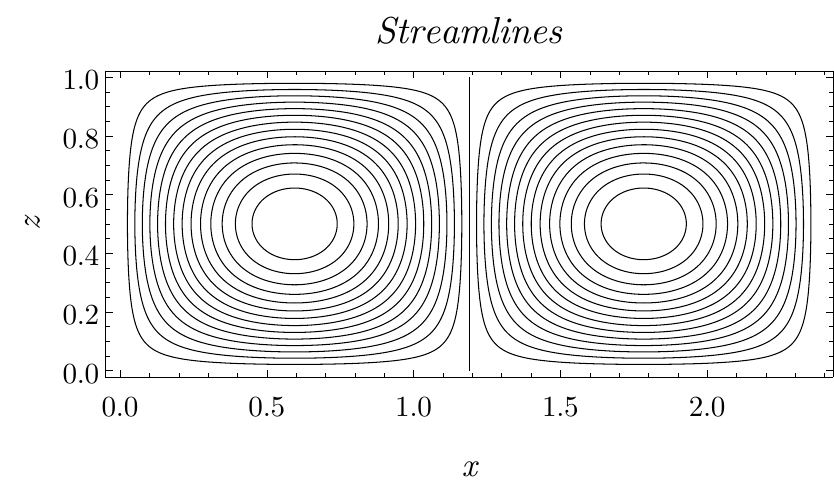}\\
\includegraphics[width=1\linewidth]{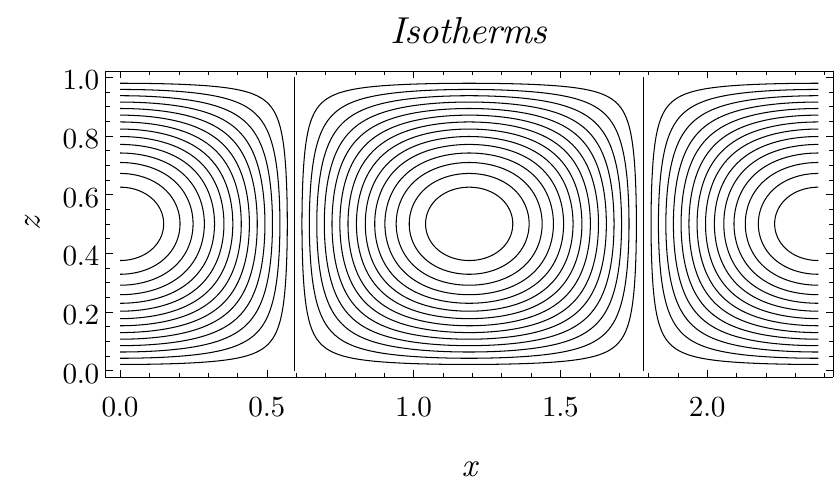}
\caption{{Streamlines and isotherms for the critical modes with $a=2$, $m=1$ and $r=0$.}}
\label{figcs}
\end{figure}

{Figure~\ref{figcs} shows the streamlines and isotherms of the perturbation normal modes at critical conditions for $a=2$, $m=1$ and $r=0$. These contour plots reveal that the cells are symmetric with respect to the midplane $z=1/2$. Compared with the isotropic Darcy--B\'enard problem, the shape of the cells is rectangular instead of square. This feature is a consequence of $k_c$ being equal to $\pi/a^{1/4}$.} 
\section{Absolute instability}
The investigation of the threshold values for the onset of absolute instability aims to study the evolution in time of a wavepacket of Fourier modes, as defined by the inverse Fourier transforms in \equa{8b}. This evaluation is carried out by employing the steepest--descent approximation \cite{barletta2019r}. By using this approximation, the long time behaviour of the integrals in \equa{8b} is determined by the sign of the real part of $\lambda$, \equa{12}. More precisely, the condition $\Re\{\lambda(k_0)\}=0$, where $k_0$ is a saddle point, {\em i.e.}, a complex zero of equation $\partial \lambda/\partial k=0$, identifies the threshold for the onset of absolute instability.  For details on this technique, we refer the reader to the book by \citet{barletta2019r}.
The necessary condition to obtain the correct saddle point for the evaluation of the threshold to absolute instability is the so--called holomorphy requirement. The real $k$ axis can be continuously deformed in the complex $k$ plane to obtain an integration path which crosses the saddle point through a direction of steepest descent. Thus, the region of the complex $k$ plane confined between the deformed integration path and the real $k$ axis cannot include a singularity of $\lambda(k)$ \cite{barletta2019r}. Equation~(\ref{12}) yields
\eqn{ 
\sigma\,\lambda=\frac{ R \left(k^2+r^2\right)}{a \left(k^2+r^2\right)+\pi ^2 m^2}\nonumber \\
\hspace{2cm}-(k^2+r^2+\pi ^2 m^2+i k Pe), \nonumber\\
\sigma\,\aderv{\lambda}{k}=\frac{2\, k\, m^2 \pi^2 R}{\left[a{ \left(k^2 + r^2\right)}+ m^2 \pi^2\right]^2} - 2 k - i Pe.
}{14}
The function $\lambda(k)$ has the singularities
\eqn{
k=\pm i\sqrt{\frac{m^2\pi ^2 +a\, r^2}{a}}.
}{17}
%
\subsection{The limiting case $Pe\to 0$}
For the limiting case of motionless basic state, $Pe\to 0$, no unstable Fourier mode can be convected away by the basic flow as the flow rate is zero. The possibility that the basic flow drives away a perturbation growing in time, so that its growth cannot be actually observed by a local measurement of the flow properties, is the physical basis of the distinction between modal and absolute instability \cite{huerre1990local}.  As a consequence, if $Pe\to 0$, we expect that the threshold value of the Darcy--Rayleigh number for the onset of absolute instability, $R_a$, coincides with the critical value, $R_c$. Therefore, among all the possible saddle points $k_0$ with $\Re\{\lambda(k_0)\}=0$, we look for that matching the critical values displayed in \equa{13}. By setting $m=1$ and $r=0$, we expect to find a purely real value of $k_0$ equal to $\pi/a^{1/4}$. For the case of $Pe\to 0$, the condition $\partial \lambda/\partial k=0$ yields the following saddle points:
\eqn{ 
k_1= 0, \quad k_2= \pm i\sqrt{\frac{\pi^2 +\pi    \sqrt{{R}}}{a}}, \nonumber  \\ 
k_3= \pm i\sqrt{\frac{\pi^2 -\pi   \sqrt{{R}}}{a}}.
}{16}
Because of the singularities reported in \equa{17}, we can exclude the saddle points $k_2$. We can exclude also $k_1$ since it does not match the critical values reported in \equa{13}. By substituting $k_3$ into the relation $\lambda(k)=0$, we obtain
\eqn{ 
R_a=\pi ^2 \left(\sqrt{a}+1\right)^2 ,
}{18}
that is what we obtained for the modal stability analysis, as displayed in \equa{13}. By substituting \equa{18} in \equa{16} we also obtain $k_3=\pi/a^{1/4}$, as expected. 

For values of the vertical permeability much higher than the values of the horizontal permeability, $K_z\gg K_h$, {\em i.e.} for $a \ll 1$, \equa{18} yields $R_a \to \pi^2$. On the other hand, for $a\gg 1$ and thus for values of horizontal permeability higher than the values of vertical permeability, $K_h\gg K_z$, \equa{18} yields $R_a \to \infty$. We can interpret the condition $a<1$ as one where the vertical flow is favoured so that the buoyancy driven cellular flow onset may happen with smaller values of $R$. The opposite behaviour is detected when $a>1$ resulting in a stabilisation of the basic state. In \fig{2}, the threshold value of $R$ for the onset of absolute instability is plotted versus $a$, for the limiting case $Pe\to 0$. As already mentioned, this threshold value, $R_a$, coincides with the critical value, $R_c$, in this case.
\begin{figure}
\centering
\includegraphics[width=0.96\linewidth]{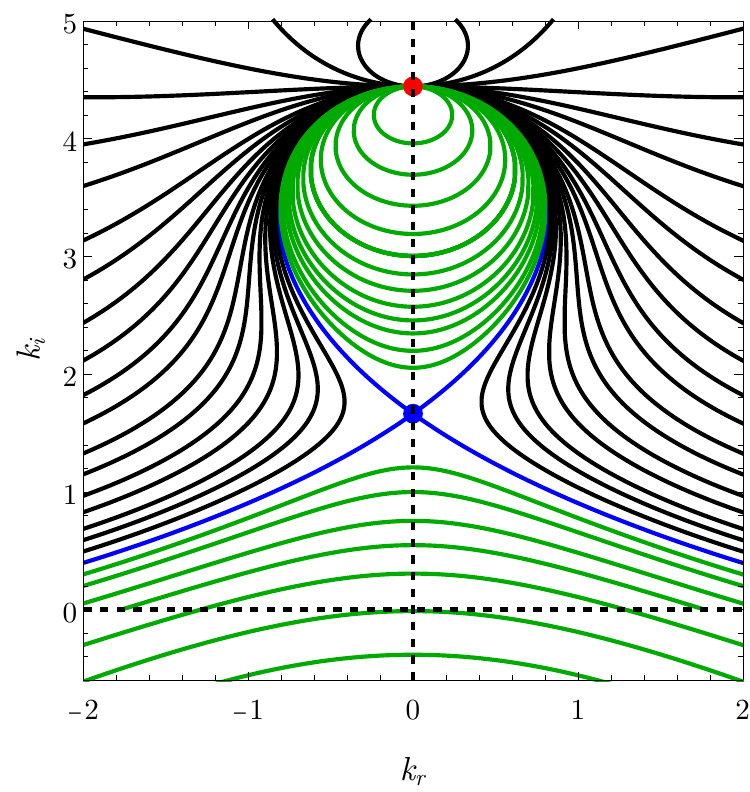}
\caption{Holomorphy requirement check in the complex $k$ plane for $a=0.5$, $m=1$, $r=0$ and $Pe=10$, at the absolute instability threshold $R=R_a = 29.2327$. The blue dot denotes the saddle point, while the red dot denotes the singularity of $\lambda(k)$. The blue lines identify the $\Re\{\lambda\}=0$ condition, while the black (green) lines are drawn for different positive (negative) values of $\Re\{\lambda\}$. 
}
\label{figcheck}
\end{figure}

\begin{figure}
\centering
\includegraphics[width=1\linewidth]{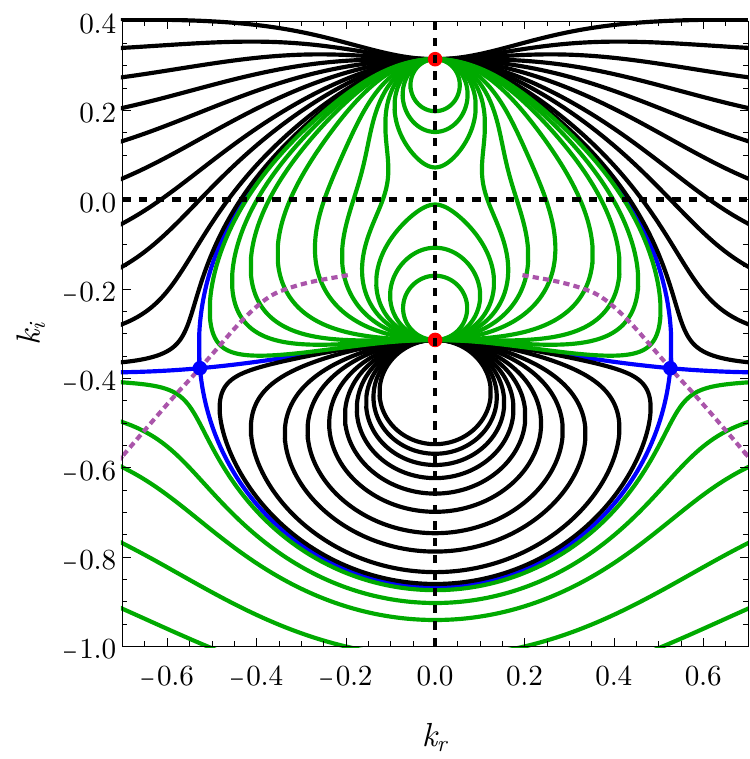}
\caption{Holomorphy requirement check in the complex $k$ plane for $m=1$, $r=0$, $Pe=10$ and $a=100$, with $R=R_a=1545.47$. The blue dots denote the saddle points, while the red dots denote the singularities of $\lambda(k)$. The blue lines identify the $\Re\{\lambda\}=0$ condition, while the black (green) lines are isolines of $\Re\{\lambda\}$ drawn for different positive (negative) values of $\Re\{\lambda\}$. The steepest descent paths are identified with purple dotted lines.
}
\label{figcheck1}
\end{figure}

\subsection{Non vanishing flow rate}
\begin{figure*}
\centering
\includegraphics[width=0.96\linewidth]{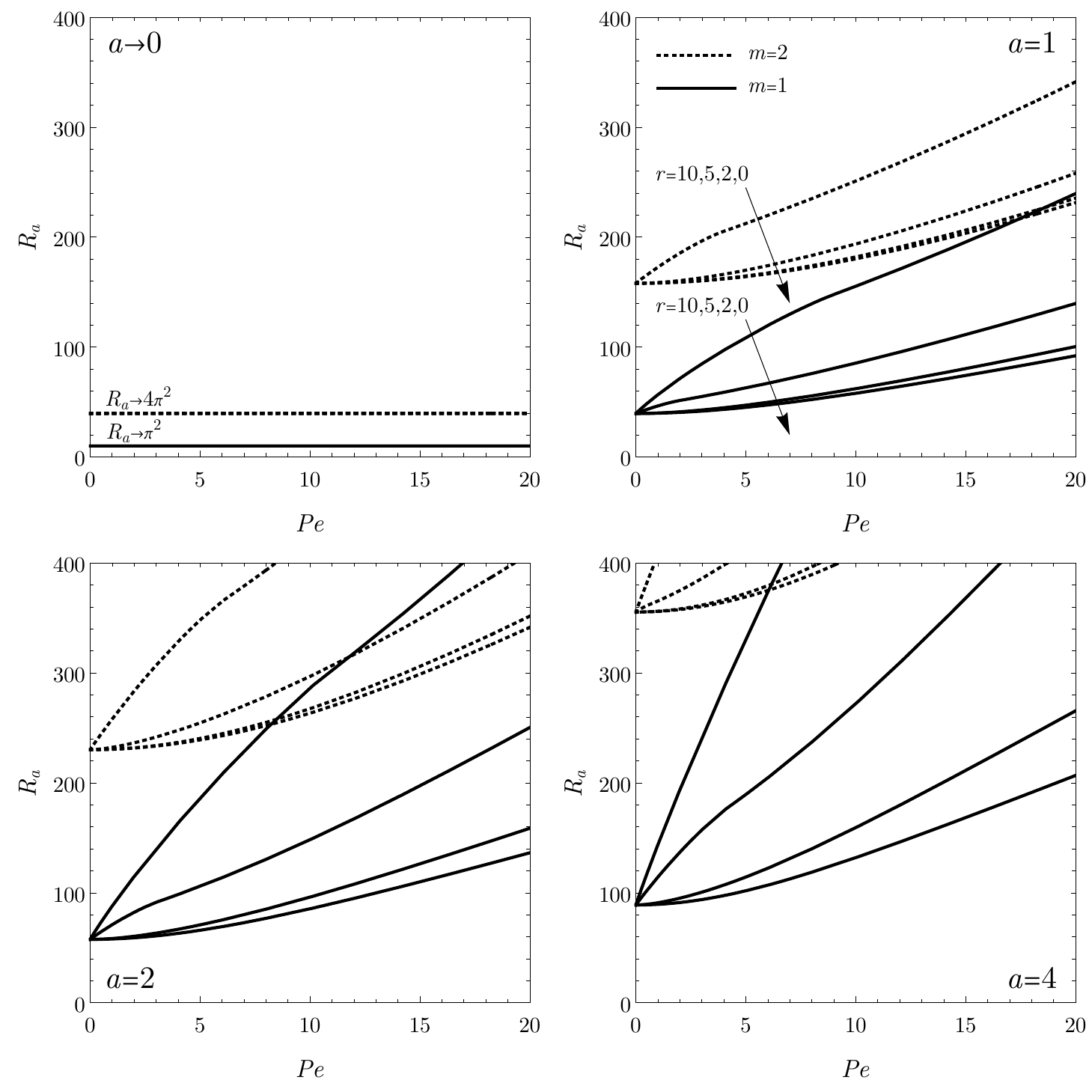}
\caption{Threshold values $R_a$ versus $Pe$ for the transition to absolute instability with different values of $m$, $r$ and $a$. The solid lines are relative to $m=1$, while the dotted lines refer to the case $m=2$. In each frame, the values of $r$ are $0,2,5,10$. The four frames correspond to $a\to 0$, $a=1$, $a=2$ and $a=4$.}
\label{figm}
\end{figure*}
When the P\'eclet number is nonzero and a flow rate is present inside the duct, looking for an analytical solution is not a convenient option. The threshold value of the Darcy--Rayleigh number, $R_a$, is  thus obtained by solving numerically the system of algebraic equations
\eqn{ 
\Re\{\lambda\}=0,\quad \Re \left \{ \aderv{\lambda}{k}\right \}=0, \quad \Im \left \{ \aderv{\lambda}{k}\right \}=0.
}{19}
%
For each fixed value of the parameter set $(a,m,Pe,r)$, there are typically multiple solutions of system~(\ref{19}). We look for the lowest value of $R_a$ constrained by the condition $R_a>R_c$ (absolute instability can arise only if at least one Fourier mode becomes unstable).  We also exclude those solutions that require a deformation of the integration path that encloses the singularities reported in \equa{17}. This occurs when the saddle point is purely imaginary, $k_r=0$, and the absolute value of the imaginary part, $k_i$, is greater than the absolute value of $k$ evaluated by employing \equa{17}. In these conditions, the singularity is included inside the deformed integration path and, hence, such a solution has to be rejected. Furthermore, some solutions have to be excluded when the steepest--descent path crosses one of the singularities~(\ref{17}). Such a behaviour is illustrated in \fig{check} for $a=0.5$, $m=1$, $r=0$, $Pe=10$ at the absolute instability threshold $R=R_a = 29.2327$. This figure displays the singularity with a red dot and the saddle point with a blue dot. The saddle point is located at the intersection of the two blue lines which identify the locus $\Re\{\lambda\}=0$. The green solid lines are isolines of $\Re\{\lambda\}$ with negative values, while the black solid lines are isolines of $\Re\{\lambda\}$ with positive values. Thus, the paths of steepest descent departing from the saddle point coincide with the axis $k_r=0$, so that the upward steepest-descent path crosses the singularity. Such a situation means that the holomorphy requirement is not satisfied.
\begin{figure}
\centering
\includegraphics[width=1\linewidth]{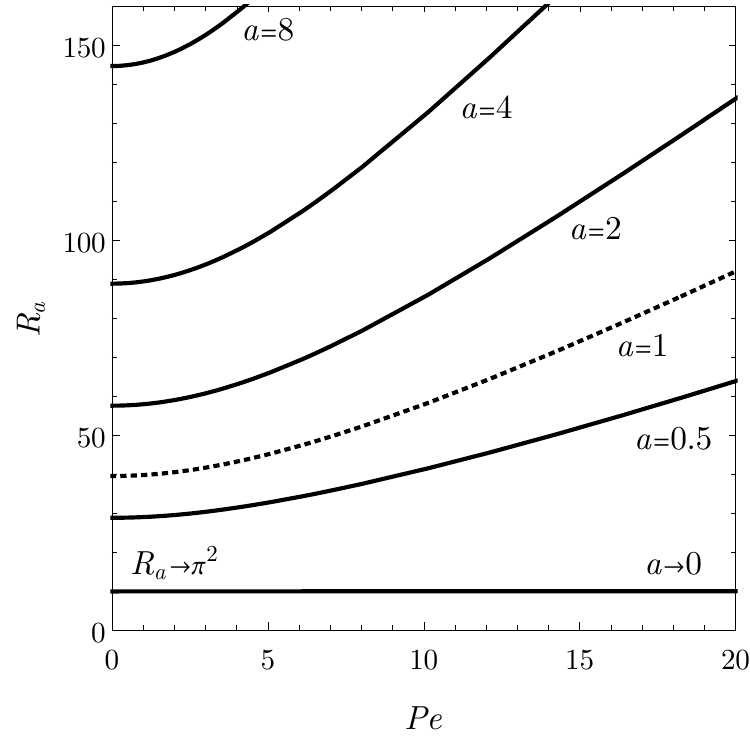}
\caption{Values of $R_a$ as a function of $Pe$ for different values of $a$. The lines are drawn for the most unstable set of the parameters $(m,r)=(1,0)$. The dotted line refer to the isotropic case $a=1$.}
\label{figRa}
\end{figure}

We present another check of the holomorphy requirement in \fig{check1}. It is a case characterised by $m=1$, $r=0$, $Pe=10$ and by $a=100$. The value of $R$ is the threshold to absolute instability, which is $R_a=1545.47$.
Figure~\ref{figcheck1} displays the singularities of $\lambda(k)$ as red dots $(k= \pm\,i\, \pi/\sqrt{a} =\pm\, i\,\pi/10)$ and the saddle points as blue dots.  The blue, green and black lines have the same meaning as in \fig{check}. The steepest descent paths are denoted as purple dotted lines. Both frames of \fig{check1} report cases where no singularity is to be encircled within the deformed integration paths locally coincident with the steepest--descent paths. Thus, the holomorphy check is to be considered as satisfied.

Figure~\ref{figm} displays $R_a$ versus $Pe$ for different values of $m$, $r$ and $a$. The solid lines denote the modes with $m=1$, while the dotted lines are relative to $m=2$. Solid and dotted lines are drawn for different values of the parameter $r$, namely $r=0,2,5,10$.  Each frame is relative to a different $a$: $a\to 0$, $a=1$, $a=2$ and $a=4$. We can conclude that $R_a$ is a monotonically increasing function of both $m$ and $r$. Thus, the transition to absolute instability happens with the lowest possible values of $m$ and $r$, that is $m=1$ and $r=0$. 

The most unstable branches of $R_a$ versus  $Pe$ (with $m=1$ and $r=0$) are plotted in \fig{Ra}, for different values of the parameter $a$. For the asymptotic case $a \to 0$, \fig{Ra} shows that $R_a$ is equal to $\pi^2$ for every $Pe$. The dotted line refers to the isotropic case, $a=1$, and it matches the results obtained by \citet{barletta2019r}. We point out that $R_a$ increases with $a$, for every $Pe$. This means that the anisotropic permeability of the porous medium implies a destabilisation when $a<1$ and a stabilisation when $a>1$. In fact, the destabilising effect of the anisotropy emerges when the permeability in the vertical direction is larger than in the horizontal direction, so that the vertical buoyant flow is favoured. On the other hand, for every fixed $a$, the threshold $R_a$ is a monotonic increasing function of $Pe$, exactly as it happens for an isotropic medium \cite{barletta2019r}. Furthermore, one may note that the dependence on $Pe$ becomes weaker and weaker as $a$ decreases. As already pointed out, in the limit $a\to 0$, $R_a$ becomes independent of $Pe$.
 
\section{Conclusions}
The linear instability of the fluid flow in a saturated horizontal porous duct heated from below has been investigated. Both the onset of the modal instability and its transition to the absolute instability have been studied. The duct has a rectangular cross--section. A basic horizontal throughflow occurs with a dimensionless velocity  equal to the P\'eclet number, $Pe$. The porous duct considered in the analysis is anisotropic: two different values of permeability, one for the horizontal directions and one for the vertical direction, are assumed. The effect of the anisotropy on the onset of absolute instability has been assessed. The governing parameter that defines the threshold value for the onset of convection is the Darcy--Rayleigh number, $R$. An exhaustive analysis of the threshold conditions for the onset of modal and absolute instability has been carried out leading to the following conclusions:
\begin{itemize}
\item As for the Prats problem, the threshold for the onset of modal instability is not affected by the presence of a horizontal throughflow: the critical values for the onset of modal instability match those reported in the literature for the Horton--Rogers--Lapwood problem within an anisotropic porous medium.
\item The principle of exchange of stabilities holds for the reference frame comoving with the basic throughflow.
\item The threshold value of the Darcy--Rayleigh number for the transition to absolute instability is a monotonic increasing function of the P\'eclet number. Thus, the basic throughflow turns out to display a stabilising effect when it comes to absolute instability.
\item In the transition to absolute instability, the threshold value of the Darcy--Rayleigh number, $R_a$, is a monotonic increasing function of the ratio between the value of the permeability for the horizontal plane and the value of the permeability for the vertical direction. In other words, the vertical permeability has a destabilising  effect on the basic state while the horizontal permeability has a stabilising effect.
\item The lowest possible value of the Darcy--Rayleigh number for the onset of absolute instability is $R_a\to \pi^2$, and it is attained when the ratio between the horizontal permeability and the vertical permeability tends to zero.
\end{itemize}

The analysis carried out in this paper accounts for the simplest situation where anisotropy emerges. There are more general cases where there can be two different horizontal permeabilities, or the three principal axes of the permeability tensor can be inclined to the horizontal. In such cases, there are more dimensionless parameters governing the anisotropy and, hence, the transition to absolute instability. Despite the increased difficulty, such a generalised scenario can be quite interesting for practical applications regarding, especially, the dynamics of hot groundwater and it will be definitely an opportunity for future research in this field.

\begin{acknowledgments}
The authors acknowledge the financial support from Grant No. PRIN 2017F7KZWS provided by the Italian Ministero dell'Istruzione, dell'Universit\`a e della Ricerca.
\end{acknowledgments}

\section*{Author Declarations}
\subsection*{Conflict of Interest}
The authors have no conflicts of interest to disclose.
\section*{Data Availability Statement}
The data that support the findings of this study are available within the article.
\nocite{*}
\bibliography{biblio}

\end{document}